\documentclass[aps,prd,onecolumn,groupedaddress,nofootinbib,amssymb]{revtex4}

\usepackage[latin1]{inputenc}
\usepackage{graphicx}
\usepackage[english]{babel}

\providecommand{\boldsymbol}[1]{\mbox{\boldmath $#1$}}

\begin{document}

\title{Gravitational waves from stellar encounters}

\author{Salvatore Capozziello$^1$, Mariafelicia De Laurentis$^2$}

\affiliation{\it $^1$Dipartimento di Scienze Fisiche, Università
di Napoli {}`` Federico II'', INFN Sez. di Napoli, Compl. Univ. di
Monte S. Angelo, Edificio G, Via Cinthia, I-80126, Napoli, Italy\\
$^2$Dipartimento di Fisica, Politecnico di Torino and INFN Sez. di
Torino, Corso Duca degli Abruzzi 24, I-10129 Torino, Italy}

\date{\today}
\begin{abstract}
The emission of gravitational waves from a system of massive
objects interacting on elliptical, hyperbolic and parabolic orbits
is studied in the quadrupole approximation. Analytical expressions
are then derived for the gravitational wave luminosity, the total
energy output and gravitational radiation amplitude. A crude
estimate of the expected number of events towards peculiar targets
(i.e. globular clusters) is also given. In particular, the rate of
events per year is obtained for the dense stellar cluster at the
Galactic Center.
\end{abstract}

%\pacs{95.85.Sz}
\maketitle


\section{\label{intro} Introduction }
Gravitational-wave (GW) science has entered a new era.
Experimentally, several GW ground-based-laser-interferometer
detectors ($10^{-1}kHz$) have been built in the United States
(LIGO) \cite{Abra}, Europe (VIRGO and GEO) \cite{Caron,Luck} and
Japan (TAMA) \cite{Ando}, and are now taking data at design
sensitivity. Advanced optical configurations capable of reaching
sensitivities slightly above and even below the so-called
standard-quantum-limit for a free test-particle, have been
designed for second \cite{Meers} and third generation
\cite{Braginsky} GW detectors. A laser-interferometer space
antenna (LISA) \cite{LISA} ($10^{-4}\sim10^{-2} Hz$) might fly
within the next decade. Resonant-bar detectors ($ \sim 1
kHz$)\cite{Astone} are improving more and more their sensitivity,
broadening their frequency band. At much lower frequencies,
$\sim10^{-17}Hz$, future cosmic microwave background (CMB) probes
are devoted to detect GWs by measuring the CMB polarization \cite{
Kamionkowski}. Millisecond pulsar timing can set interesting upper
limits \cite{Jenet} in the frequency range $10^{-9}\sim 10^{-8}
Hz$. In this frequency range, the large number of millisecond
pulsars which will be detectable with the square kilometer array
\cite{skatelescope}, would provide an ensemble of clocks that can
be used as multiple arms of a GW detector.

From a theoretical point of view, recent years have been
characterized by numerous major advances due, essentially, to the
development of numerical gravity. Concerning the most promising
sources to be detected, the GW generation problem has improved
significantly in relation to the dynamics of binary and multiple
systems of compact objects as neutron stars and black holes.

Besides, the problem of non-geodesic motion of particles in curved
spacetime has been developed considering  the emission of GWs
\cite{Poisson,Mino}. Solving this problem is of considerable
importance in order to predict  the accurate waveforms of GWs
emitted by extreme mass-ratio binaries, which are among the most
promising sources for LISA \cite{Finn}. To  this aim, searching
for  criteria to classify   the ways in which sources collide is
of fundamental importance.  A first rough  criterion can be the
classification of stellar encounters in {\it collisional} as in
the globular clusters and in {\it collisionless}  as in the
galaxies \cite{binney}.  A fundamental parameter is the richness
and the density of the stellar system  and so, obviously, we
expect a large production of GWs in rich and dense systems.

Systems with these features are the globular clusters and the
galaxy centers. In particular, one can take into account  the
stars (early-type and late-type) which are around our Galactic
Center, e.g. Sagittarius $A^{*}$ ($Sgr A^{*}$) which could be very
interesting targets for the above mentioned ground-based and
space-based detectors.

In recent years, detailed information has been achieved for
kinematics and dynamics of stars moving in the gravitational field
of such a central object. The statistical properties of spatial
and kinematical distributions are of particular interest (see e.g.
\cite{Genzel,Sellgreen,CapozIov}). Using them, it is possible to
give a quite accurate estimate of the mass and the size of the
central object: we have  $(2.61\pm0.76)\times10^6M_{\odot})$
concentrated within a radius of $0.016 pc$ (about $30$
light-days)\cite{Ghez,Thatte}. More precisely, in \cite{Ghez}, it
is described a campaign of observations where velocity
measurements in the central $arcsec^{2}$ are extremely accurate.
Then  from this bulk of data, considering  a field of resolved
stars whose proper motions are accurately known, one can classify
orbital motions and deduce, in principle,  the rate of production
of GWs according to the different types of orbits. This motivates
this paper in which, by a classification of orbits in accordance
with the conditions of motion, we want to calculate the GW
luminosity for the different types of stellar encounters. A
similar approach has been developed in \cite{achille} but, in that
case, only hyperbolic trajectories have been considered.

Following the method outlined in  \cite{pm1,pm2}, we investigate
the GW emission by  binary systems  in the quadrupole
approximation considering bounded   (circular or elliptical) and
unbounded  (parabolic or hyperbolic) orbits. Obviously, the main
parameter is the approaching energy of the stars in the system
(see also \cite{schutz} and references therein).  We expect that
gravitational waves are emitted with a "peculiar" signature
related to the encounter-type: such a signature has to be a
"burst" wave-form with a maximum in correspondence of the
periastron distance. The problem of {\it bremsstrahlung}-like
gravitational wave emission has been studied in detail by Kovacs
and Thorne \cite{kt} by considering stars interacting on unbounded
and bounded orbits. In this paper, we face this problem discussing
in detail the dynamics of such a phenomenon which could greatly
improve the statistics of possible GW sources.

The paper is organized as follows: in Sec. II, the main features
of stellar encounters and orbit classification are reviewed. Sec.
III is devoted to the emission and luminosity of GWs from binary
systems in the different kinds of orbits  assuming the quadrupole
approximation. A discussion of the  wave-form dependence from the
orbital parameters is given in Sec. IV. In Sec. V, we derive the
expected rate of events assuming the Galactic Center as a target.
Section VI is devoted to concluding remarks.

\section{Orbits in stellar encounters}

Let us take into account the Newtonian theory of orbits since
stellar systems, also if at high densities and constituted by
compact objects, can be usually assumed in Newtonian regime. We
give here a self-contained summary of the well-known orbital types
in order to achieve below a clear classification of the possible
GW emissions. We refer to the text books \cite{binney,landau} for
a detailed discussion.

A mass $m_1$ is moving in the gravitational potential $\Phi$
generated by a second mass $m_2$. The  vector radius and the polar
angle depend on time as a consequence of the star motion, i.e.
$\textbf{r}=\textbf{r}(t)$ and $\phi=\phi(t)$. With this choice,
the velocity $\textbf{v}$ of the mass $m_1$ can be parameterized
as
\begin{equation}
\textbf{v}=v_r\widehat{r}+v_{\phi}\widehat{\phi}~,
\end{equation}
where the radial and tangent components of the velocity are,
respectively,
\begin{equation}
v_r=\frac{dr}{dt} ~~~~~~~~v_{\phi}=r \frac{d\phi}{dt}~.
\end{equation}
In this case, the total energy and the angular momentum, read out
\begin{equation}
\frac{1}{2}{\mu\left(\frac{dr}{dt}\right)}^{2}+\frac{\mathbb{L}^{2}}{2\mu r^{2}}-\frac{\gamma}{r}=E
\label{eq:energia}\end{equation}
and
\begin{equation}
L=r^2\frac{d\phi}{dt} \label{eq:momang1},
\end{equation}
respectively, where $\mu=\frac{m_1m_2}{m_1+m_2}$ is the reduced
mass of the system and $\gamma=Gm_1m_2$.

We can split the kinetic energy into two terms where, due to the
conservation of angular momentum, the second one is a function of
$r$ only. An effective potential energy $V_{eff}$,
\begin{equation}
V_{eff}=\frac{\mathbb{L}^{2}}{2\mu r^{2}}-\frac{\gamma}{r}\label{eq:energpot}
\end{equation}
is immediately defined. The first  term corresponds to a repulsive
force, called the angular momentum barrier. The second term is the
gravitational attraction. The interplay between attraction and
repulsion is such that the effective potential energy has a
minimum. Indeed, differentiating with respect to $r$ one finds
that the minimum lies at $r_0=\frac{L^{2}}{\gamma\mu}$ and that
\begin{equation}
V_{eff}^{min}=-\frac{\mu\gamma^{2}}{2L^{2}}\label{eq:enrgpotmin}\,.\end{equation}
Therefore, since the radial part of kinetic energy,
\begin{equation}
K_{r}=\frac{1}{2}\mu\left(\frac{dr}{dt}\right)^{2}
\end{equation},
is non-negative, the total energy must be not less than
$V_{eff}^{min}$, i.e.
\begin{equation}
E\geq
E_{min}=-\frac{\mu\gamma^{2}}{2L^{2}}\label{eq:emin}\,.\end{equation}
The equal sign corresponds to the radial motion. For
$E_{min}<E<0$, the trajectory lies between a smallest value
$r_{min}$ and greatest value $r_{max}$ which can be found from the
condition $E=V_{eff}$, i.e.
\begin{equation}
r_{\{min,max\}}=-\frac{\gamma}{2E}\pm\sqrt{\left(\frac{\gamma}{2E}\right)^{2}+\frac{L^{2}}{2\mu
E}}\label{eq:rminmax}\end{equation} where the upper (lower) sign
corresponds to $r_{max}$ ($r_{min}$). Only for $E>0$, the upper
sign gives an acceptable value; the second root is negative and
must be rejected.

Let us now proceed in solving the  differential equations
(\ref{eq:energia}) and (\ref{eq:momang1}). We have

\begin{equation}
\frac{dr}{dt}=\frac{dr}{d\phi}\frac{d\phi}{dt}=\frac{L}{\mu r^{2}}\frac{dr}{d\phi}=
-\frac{L}{\mu}\frac{d}{d\phi}\left(\frac{1}{r}\right)\label{eq:diff}\end{equation}

and  defining, as standard,  the auxiliary variable $u=1/r$,  Eq.
(\ref{eq:energia}) takes the form
\begin{equation}
u'^{2}+u^{2}-\frac{2\gamma\mu}{L^{2}}u=\frac{2\mu E}{L^{2}}\label{eq:diffenerg}\end{equation}

where $u'=du/d\phi$ and we have divided by $L^{2}/2\mu$. Differentiating
with respect to $\phi$, we get

\begin{equation}
u'\left(u''+u-\frac{\gamma\mu}{L^{2}}\right)=0\end{equation}

hence either $u'=0$, corresponding to the circular motion, or

\begin{equation}
u''+u=\frac{\gamma\mu}{L^{2}}\label{eq:moto}\end{equation}

which has the solution
\begin{equation}
u=\frac{\gamma\mu}{L^{2}}+C\cos\left(\phi+\alpha\right)
\end{equation}

or,  reverting the variable,

\begin{equation}
r=\left[\frac{\gamma\mu}{L^{2}}+C\cos\left(\phi+\alpha\right)\right]^{-1}\label{eq:solution}\end{equation}

which is the canonical form of conic sections in polar coordinates
\cite{smart}. The constant $C$ and $\alpha$ are two integration
constants of the second order differential equation
(\ref{eq:moto}). The solution (\ref{eq:solution}) must satisfy the
first order differential equation (\ref{eq:diffenerg}).
Substituting (\ref{eq:solution}) into (\ref{eq:diffenerg}) we
find, after a little algebra,

\begin{equation}
C^{2}=\frac{2\mu E}{L^{2}}+\left(\frac{\gamma\mu}{L^{2}}\right)^{2}\label{eq:C}
\end{equation}

and therefore, taking account of Eq. (\ref{eq:emin}), we get
$C^{2}\geq 0$. This implies the four kinds of orbits given in
Table I.

%%%%%%%%
\begin{table}
\[
\]
\begin{center}
\begin{tabular}{|c|c|c|cl}
\hline $C=0$& $E=E_{min}$& circular orbits\tabularnewline \hline
$0<\left|C\right|<\frac{\gamma\mu}{L^{2}}$& $E_{min}<E<0$&
elliptic orbits\tabularnewline \hline
$\left|C\right|=\frac{\gamma\mu}{L^{2}}$& $E=0$& parabolic
orbits\tabularnewline \hline
$\left|C\right|>\frac{\gamma\mu}{L^{2}}$& $E>0$,& hyperbolic
orbits\tabularnewline \hline
\end{tabular}
\end{center}\[
\]
\caption{Orbits in Newtonian regime classified by the approaching
energy.}
\end{table}
%%%%%%%%%%

\subsection{Circular Orbits}
Circular motion corresponds to the condition $u'=0$ by which one
find $r_{0}=L^{2}/\mu\gamma$ where $V_{eff}$ has its minimum. We
also note that the expression for $r_{0}$ together with
Eq.(\ref{eq:emin}) gives\begin{equation}
r_{0}=-\frac{\gamma}{2E_{min}}\label{eq:rzero}\end{equation}

Thus the two bodies move in concentric circles with radii, inversely
proportional to their masses and are always in opposition.

\subsection{Elliptical  Orbits}

For $0<\left|C\right|<\mu\gamma/L^{2}$, $r$ remains finite for all
values of $\phi$. Since  $r(\phi+2\pi)=r(\phi)$, the trajectory is
closed and it is an ellipse. If one chooses $\alpha=0$,  the major
axis of the ellipse corresponds to $\phi=0$. We get

\begin{equation}
r_{\left|\phi=0\right.}=r_{min}=\left[\frac{\gamma\mu}{L^{2}}+C\right]^{-1}\label{eq:rphi}\end{equation}

and

\begin{equation}
r_{\left|\phi=\pi\right.}=r_{max}=\left[\frac{\gamma\mu}{L^{2}}-C\right]^{-1}\label{eq:rpi}\end{equation}

and since $r_{max}+r_{min}=2a$, where $a$ is the semi-major axis
of the ellipse, one obtains

\[
a=r_{\left|\phi=0\right.}=r_{min}=\frac{\gamma\mu}{L^{2}}\left[\left(\frac{\gamma\mu}{L^{2}}\right)^{2}+C^{2}\right]^{-1}\]

$C$ can be eliminated from the latter equation and Eq.(
\ref{eq:C}) and then

\begin{equation}
a=-\frac{\gamma}{2E}\label{eq:a}\end{equation}

Furthermore, if we denote the distance
$r_{\left|\phi=\pi/2\right.}$ by $l$, the so-called {\it
semi-latus rectum} or the parameter of the ellipse,  we get

\begin{equation}
l=\frac{L^{2}}{\gamma\mu}\label{eq:latusrectum}\end{equation}

and hence the equation of the trajectory

\begin{equation}
r=\frac{l}{1+\epsilon\cos\phi}\label{eq:traiettoria}\end{equation}

where $\epsilon=\sqrt{\frac{1-l}{a}}$ is the eccentricity of the
ellipse.
 \subsection{Parabolic and Hyperbolic Orbits}

These solutions can be dealt together. They correspond to $E\geq
0$ which is the condition to obtain unbounded orbits.
Equivalently, one has $\left|C\right|\geq\gamma\mu/L^{2}$.

The trajectory is
\begin{equation}
r=l\left(1+\epsilon\cos\phi\right)^{-1}\label{eq:traie}\end{equation}

where $\epsilon\geq1$. The equal sign corresponds to $E=0$ .
Therefore, in order to ensure positivity of $r$, the polar angle
$\phi$ has to be restricted to the range given by

\begin{equation}
1+\epsilon\cos\phi>0 \label{eq:cosphi}\end{equation}

This means $\cos\phi>-1$, i.e. $\phi\in(-\pi,\pi)$ and the
trajectory is  not closed any more. For $\phi\rightarrow\pm\pi$,
we have $r\rightarrow\infty$. The curve (\ref{eq:traie}), with
$\epsilon=1$, is a parabola. For  $\epsilon>1$, the allowed
interval of polar angles is smaller than $\phi\in(-\pi,\pi)$, and
the trajectory is a hyperbola. Such  trajectories correspond to
non-returning objects.

\section{Gravitational wave luminosity in the quadrupole
approximation}

At this point, considering the orbit equations,  we want to
classify the gravitational radiation for the different stellar
encounters. To this aim, let us start with a short review of the
quadrupole approximation for GW radiation. We add this discussion
for the sake of completeness, but we send the Reader to the
References \cite{gravitation,shapiro,maggiore,thorne} for a
detailed exposition.

The Einstein field equations give a description of how the
curvature of space-time is related to the energy-momentum
distribution. In the weak field approximation, moving massive
objects produce gravitational waves which propagate in the vacuum
with the speed of light. In this approximation, we have

\begin{equation}
g_{\mu\nu}=\delta_{\mu\nu}+\kappa h_{\mu\nu},\qquad\left(\left|h_{\mu\nu}\right|<<1\right),
\end{equation}
whit $\kappa$ the gravitational coupling. The field equations are

\begin{equation}
\square\bar{h}_{\mu\nu}=-\frac{1}{2}\kappa T_{\mu\nu}\label{eq:h}\end{equation}
where

\begin{equation}
\bar{h}_{\mu\nu}=h_{\mu\nu}-\frac{1}{2}\delta_{\mu\nu}h_{\lambda\lambda}\,,\end{equation}

and $T_{\mu\nu}$ is the total stress-momentum-energy tensor of the
source, including the gravitational stresses.

A plane GW can be written as

\begin{equation}
\label{wave} \bar{h}_{\mu\nu}=h_{\mu\nu}=he_{\mu\nu}\cos(\omega
t-\boldsymbol{\mathbf{k}\cdot\mathbf{x}})
\end{equation}

where $h$ is the amplitude, $\omega$ the frequency, $k$ the wave
number and $e_{\mu\nu}$ is a unit polarization tensor, obeying the
conditions

\begin{equation}
e_{\mu\nu}=e_{\nu\mu},\qquad e_{\mu\mu}=0,\qquad e_{\mu\nu}e_{\mu\nu}=1.\end{equation}

Let us assume a  gauge in which $e_{\mu\nu}$ is space-like and
transverse; thus, a wave travelling in the \emph{z} direction has
two possible independent polarizations:

\begin{equation}
e_{1}=\frac{1}{\sqrt{2}}(\hat{x}\hat{x}-\hat{y}\hat{y})\qquad e_{2}=\frac{1}{\sqrt{2}}(\hat{x}\hat{y}-\hat{y}\hat{x}).\end{equation}

One can now search for wave solutions of (\ref{eq:h}) from a
system of masses undergoing arbitrary motions, and then obtain the
power radiated. The result, assuming the source dimensions very
small with respect to the wavelengths (quadrupole approximation
\cite{landau}), is that the power ${\displaystyle
\frac{dE}{d\Omega}}$ radiated in a solid angle $\Omega$ with
polarization $e_{ij}$ is

\[
\]
\begin{equation}
\frac{dE}{d\Omega}=\frac{G}{8\pi c^{5}}\left(\frac{d^{3}Q_{ij}}{dt^{3}}e_{ij}\right)^{2}\label{eq:P}\end{equation}
where $Q_{ij}$ is the  quadrupole mass tensor
\begin{equation}
Q_{ij}=\sum _a
m_a(3x_a^ix_a^j-\delta_{ij}r_a^2)~,\label{qmasstensor}
\end{equation}
$G$ being the Newton constant,  $r_a$  the modulus of the vector
radius of the $a-th$ particle and the sum running over all masses
$m_{a}$ in the system. It has to be noted that the result is
independent of the kind of stresses which are present into the
dynamics. If one sums (\ref{eq:P}) over the two allowed
polarizations, one obtains
\begin{eqnarray}
\sum_{pol}\frac{dE}{d\Omega} & = & \frac{G}{8\pi c^{5}}\left[\frac{d^{3}Q_{ij}}{dt^{3}}\frac{d^{3}Q_{ij}}{dt^{3}}-2n_{i}\frac{d^{3}Q_{ij}}{dt^{3}}n_{k}\frac{d^{3}Q_{kj}}{dt^{3}}-\frac{1}{2}\left(\frac{d^{3}Q_{ii}}{dt^{3}}\right)^{2}\right.\nonumber \\
 &  & \left.+\frac{1}{2}\left(n_{i}n_{j}\frac{d^{3}Q_{ij}}{dt^{3}}\right)^{2}+\frac{d^{3}Q_{ii}}{dt^{3}}n_{j}n_{k}\frac{d^{3}Q_{jk}}{dt^{3}}\right]\label{eq:sommatoria}\end{eqnarray}

where $\hat{n}$ is the unit vector in the  radiation direction.
The total radiation rate is obtained by integrating
(\ref{eq:sommatoria}) over all directions of emission; the result
is

\begin{equation}
\frac{dE}{dt}=-\frac{G\left\langle Q_{ij}^{(3)}Q^{(3)ij}\right\rangle }{45c^{5}}\label{eq:dEdt}\end{equation}

where the index (3) represents the differentiation with respect to
time, the symbol $<>$ indicates that the quantity is averaged over
several wavelengths. This crucial point is linked with the
difficulties of localizing gravitational energy so the right hand
side of Eq. (\ref{eq:dEdt}) cannot be viewed as an instantaneous
quantity. This problem has been already faced for circular and
elliptical orbits in \cite{Finn,bla,buonanno}. For hyperbolic and
parabolic  orbits, it is crucial to estimate the quantity in the
right hand side of Eq.(\ref{eq:dEdt}) in the zone where stars are
slightly changing their trajectories, that means at peri-astron,
while we expect no emission in asymptotic regime of stars
approaching to or going away from this region. For a detailed
discussion in the hyperbolic case, see \cite{achille}.

With this formalism,  it is possible to estimate the amount of
energy emitted in the form of GWs from a system of massive objects
interacting among them \cite{pm1,pm2}. In this case, the
components of the quadrupole mass tensor in the equatorial plane
($\theta=\pi/2$) are
\begin{equation}
\begin{array}{lll}
Q_{xx}=\mu r^2(3\cos{^2\phi}-1)~,\\ \\
Q_{yy}=\mu r^2(3\sin{^2\phi}-1)~,\\ \\
Q_{zz}=-\mu r^2~,\\ \\
Q_{xz}=Q_{zx}=0~,\\ \\
Q_{yz}=Q_{zy}=0~,\\ \\
Q_{xy}=Q_{yx}=3\mu r^2 \cos\phi \sin\phi~,
\end{array}\label{eq:quadrupoli}
\end{equation}
where the masses $m_{1}$ and $m_{2}$ have  polar coordinates
$\{r_{i}\cos\theta\cos\phi,\; r_{i}\cos\theta\sin\phi,\:
r_{i}\sin\theta\}$ with $i=1,2$. The origin of the motions is
taken at the center of mass. Such components can be differentiated
with respect to the time as in Eq.(\ref{eq:dEdt}). In doing so, we
can  use some useful relations derived in the previous Section.

\subsection{GW luminosity from circular and elliptical orbits}

Using  Eq:(\ref{eq:traiettoria}), let us derive  the angular
velocity equation
\begin{equation}
\dot{\phi}=\frac{\sqrt{G l (m_{1}+m_{2})} (\epsilon  \cos\phi+1)^2}{l^2}
\label{eq:angularvelo}
\end{equation}

and then, from (\ref{eq:quadrupoli}), the  quadrupolar components
for the elliptical orbits

\begin{equation}
\frac{d^{3}Q_{xx}}{dt^{3}}=\beta(24 \cos\phi+\epsilon  (9 \cos2 \phi)+11)) \sin
   \phi
\end{equation}

\begin{equation}
\frac{d^{3}Q_{yy}}{dt^{3}}=-\beta(24 \cos\phi+\epsilon  (13+9 \cos2 \phi)) \sin\phi)
\end{equation}

\begin{equation}
\frac{d^{3}Q_{zz}}{dt^{3}}=-2\beta \epsilon \sin\phi
\end{equation}

\begin{equation}
\frac{d^{3}Q_{xy}}{dt^{3}}=\beta
(24 \cos\phi+\epsilon  (11+9 \cos2 \phi)) \sin\phi)
\end{equation}

where
\begin{equation}
\beta=\frac{G l (m_{1}+m_{2}))^{3/2} \mu  (\epsilon
\cos\phi+1)^2}{l^4}\,.
\end{equation}
Being
\begin{eqnarray*}
 Q_{ij}^{(3)}Q^{(3)ij}=\frac{G^3}{l^5}\left[ (m_{1}+m_{2})^3 \mu ^2 (1+\epsilon  \cos\phi)^4\right.\\
\left(415 \epsilon ^2+3 (8 \cos\phi+3 \epsilon  \cos2 \phi) \right.\\
\left.(72 \cos\phi+\epsilon  (70+27 \cos2 \phi)))
\sin\phi^{2}\right]
 \end{eqnarray*}
the total power radiated is  given by
\begin{eqnarray*}
\frac{dE}{dt}=\frac{G^3}{45c^5l^5}f(\phi)\end{eqnarray*}
where
\begin{equation}
f(\phi)=\left[ (m_{1}+m_{2})^3 \mu ^2 (1+\epsilon  \cos\phi)^4\right.\\
\left(415 \epsilon ^2+3 (8 \cos\phi+3 \epsilon  \cos2 \phi) \right.\\
\left.(72 \cos\phi+\epsilon  (70+27 \cos2 \phi)))
\sin\phi^{2}\right]
\end{equation}
The total energy emitted in the form of gravitational radiation,
during the interaction, is given by
\begin{equation}
\Delta E=\int^{\infty}_0 \left|\frac{dE}{dt}\right| dt~.
\end{equation}
From Eq.(\ref{eq:momang1}), we can adopt the angle $\phi$ as a
suitable integration variable. In this case, the energy emitted
for $\phi_1<\phi<\phi_2$ is
\begin{equation}
\Delta E(\phi_1,\phi_2)
=\frac{G^3}{45c^5l^5}\int^{\phi_2}_{\phi_1}f(\phi)~d\phi~,\label{eq:integraleenergia}
\end{equation}
and the total energy can be determined from the previous relation
in the limits $\phi_1\rightarrow 0$ and $\phi_2\rightarrow
\pi$. Thus, one has
\begin{equation}
\Delta E=\frac{G^4 \pi  (m_{1}+m_{2})^3 \mu ^2}{l^5c^5}F(\epsilon)~
\end{equation}
where $F(\epsilon)$ depends on the initial conditions only and is
given by
\begin{equation}
F(\epsilon)=\frac{ \left(13824+102448 \epsilon ^2+59412 \epsilon ^4+2549 \epsilon ^6\right)}{2880}
\end{equation}
In other words, the gravitational wave luminosity strictly depends
on the configuration and kinematics of the binary system.

\subsection{GW luminosity from parabolic and hyperbolic orbits}

Also in this case, we use   Eq:(\ref{eq:traie})  and Eq.
(\ref{eq:quadrupoli}) to calculate the quadrupolar formula for
parabolic and hyperbolic orbits. The angular velocity is

\begin{equation}
\dot{\phi}=l^2 L (\epsilon  \cos\phi+1)^2
\label{eq:angularvelo1}
\end{equation}
and the derivative are

\begin{equation}
\frac{d^{3}Q_{xx}}{dt^{3}}=\rho(24 \cos\phi+\epsilon
    (9 \cos 2 \phi+11)) \sin \phi
\end{equation}

\begin{equation}
\frac{d^{3}Q_{yy}}{dt^{3}}=-\rho(24 \cos\phi+\epsilon  (13+9 \cos2 \phi)) \sin\phi)
\end{equation}

\begin{equation}
\frac{d^{3}Q_{zz}}{dt^{3}}=-2\rho\epsilon\sin\phi
\end{equation}

\begin{equation}
\frac{d^{3}Q_{xy}}{dt^{3}}=-\frac{3}{2}\rho(\epsilon  \cos \phi+1)^2 (5 \epsilon
   \cos \phi+8 \cos 2 \phi+3 \epsilon  \cos3 \phi)
\end{equation}
where
\begin{equation}
\rho=l^4 L^3 \mu  (\epsilon  \cos\phi+1)^2\,.
\end{equation}
The  radiated power is given by
\begin{eqnarray*}
\frac{dE}{dt}=-\frac{G \rho^2 \left([314
   \epsilon ^2+(1152 \cos (\phi+187 \epsilon  \cos 2 \phi-3
   (80 \cos 3 \phi+30 \epsilon  \cos4 \phi+48 \cos 5 \phi+9 \epsilon  \cos6 \phi)) \epsilon
   -192 \cos4 \phi+576\right]}{120 c^5}
\end{eqnarray*}

then
\begin{equation}
\frac{dE}{dt}=-\frac{G l^8 L^6 \mu ^2 }{120 c^5 }f(\phi)
\end{equation}

where $f(\phi)$, in this case, is
\begin{equation}
f(\phi)= \left(314
   \epsilon ^2+(1152 \cos (\phi+187 \epsilon  \cos 2 \phi-3
   (80 \cos 3 \phi+30 \epsilon  \cos4 \phi+48 \cos 5 \phi+9 \epsilon  \cos6 \phi)) \epsilon -192 \cos4 \phi+576\right)
\end{equation}

Then using  Eq. (\ref{eq:dEdt}), the total energy emitted in the
form of gravitational radiation during the interaction as a
function of $\phi$ is given by
\begin{equation}
\Delta E(\phi_1,\phi_2)
=-\frac{G l^8 L^6 \pi  \left(1271 \epsilon ^6+24276 \epsilon ^4+34768
   \epsilon ^2+4608\right) \mu ^2}{480 c^5}
~d\phi~,\label{eq:integraleenergia1}
\end{equation}
and the total energy can be determined from the previous relation
in the limits $\phi_1\rightarrow -\pi$ and $\phi_2\rightarrow\pi$ in the parabolic case. Thus, one has
\begin{equation}
\Delta E=-\frac{G l^8 L^6 \pi \mu^2 }{480 c^5}F(\epsilon)~,
\end{equation}
where $F(\epsilon)$ depends on the initial conditions only and is
given by
\begin{equation}
F(\epsilon)= \left(1271 \epsilon ^6+24276 \epsilon ^4+34768
   \epsilon ^2+4608\right)~.
 \end{equation}
 In the hyperbolic case, we have that the total energy is determined in the limits
  ${\displaystyle \phi_1\rightarrow \frac{-3\pi}{4}}$ and ${\displaystyle \phi_2\rightarrow\frac{-3\pi}{4}}$, i.e.
 \begin{equation}
\Delta E=--\frac{G l^8 L^6\mu^2}{201600 c^5}F(\epsilon)~,
\end{equation}
where $F(\epsilon)$ depends on the initial conditions only and is
given by
\begin{eqnarray}
%\begin{array}{ll}
F(\epsilon)&=& \left[315 \pi  \left(1271 \epsilon ^6+24276
\epsilon ^4+34768 \epsilon
   ^2+4608\right)+\right.\\ & &  \left. +16 \epsilon  \left[\epsilon  \left[\epsilon  \left(926704 \sqrt{2}-7
   \epsilon  (3319 \epsilon ^2-32632 \sqrt{2} \epsilon
   +55200)\right)-383460\right]+352128 \sqrt{2}\right]\right]~.
%\end{array}
\end{eqnarray}

As above, the gravitational wave luminosity strictly depends on
the configuration and kinematics of the binary system.

\section{Gravitational wave amplitude}

Direct signatures of gravitational radiation are its amplitude and
its wave-form. In other words, the identification of a GW signal
is strictly related to the accurate selection of  the shape of
wave-forms by interferometers or any possible detection tool. Such
an achievement could give information on the nature of the GW
source, on the propagating medium, and , in principle, on the
gravitational theory producing such a radiation \cite{Dela}.

It is well known that the amplitude of GWs can be evaluated by
\begin{equation}
h^{jk}(t,R)=\frac{2G}{Rc^4}\ddot{Q}^{jk}~, \label{ampli1}
\end{equation}
$R$ being the distance between the source and the observer and
$\{j,k\}=1,2$. Let us now derive the GW amplitude in relation to
the orbital shape of the binary systems.

\subsection{GW amplitude from elliptical orbits}

\begin{figure}
\includegraphics[scale=0.6]{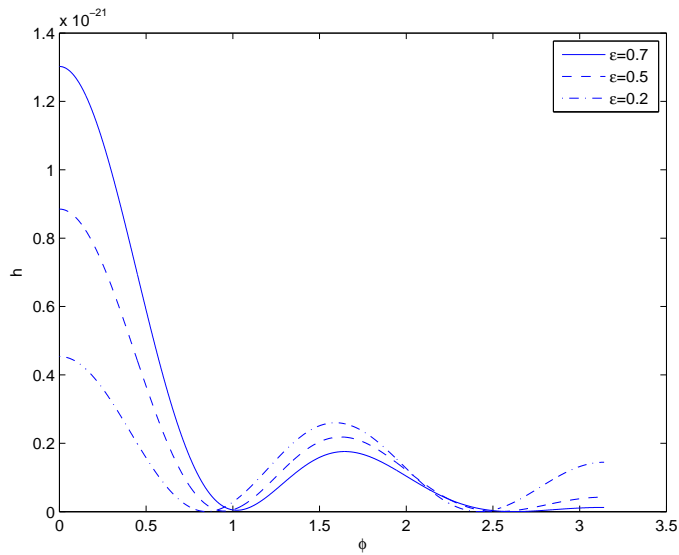}
\caption{The gravitational wave-forms from elliptical orbits shown
as  function of the polar angle $\phi$.  We have fixed
$M_{1}=M_{2}=1.4M_{\odot}$. $M_{2}$ is considered at rest while
$M_{1}$ is moving with initial velocity $v_{0}=200$ Km$s^{-1}$ and
an impact parameter $b=1$ AU. The distance of the GW source is
assumed to be $R=8$ kpc and the eccentricity is  $\epsilon=
0.2,0.5, 0.7.$ } \label{fig:ellisse800pc}
\end{figure}

Considering a binary system and the single components of
eq.(\ref{ampli1}), it is straightforward to show that
\begin{equation}
\begin{array}{llllllll}
h^{11}=-\frac{2G}{Rc^4}\frac{G (m_{1}+m_{2}) \mu  (13 \epsilon  \cos \phi+12 \cos2 \phi+\epsilon  (4 \epsilon +3 \cos3 \phi))}{2 l}~,
\\ \\
h^{22}=\frac{2G}{Rc^4}\frac{G (m_{1}+m_{2})  \mu  (17 \epsilon  \cos\phi+12 \cos2 \phi+\epsilon  (8 \epsilon +3 \cos3 \phi))}{2 l}
 ~,
\\ \\
h^{12}=h^{21}=-\frac{2G}{Rc^4}\frac{G (m_{1}+m_{2})  \mu  (13 \epsilon  \cos\phi+12 \cos2 \phi+\epsilon  (4 \epsilon +3 \cos3 \phi))}{2
l}
~,
\end{array}
\end{equation}
so that the expected strain amplitude
$h\simeq(h_{11}^2+h_{22}^2+2h_{12}^2)^{1/2}$ turns out to be

\begin{eqnarray}
h=\frac{G^3 (m_{1}+m_{2})  \mu^2}{c^4 Rl^2}
  (3 (13 \epsilon  \cos\phi+12 \cos2 \phi+\epsilon  (4 \epsilon
+3
\cos3 \phi))^2+(17 \epsilon  \cos\phi+
12 \cos2 \phi+\epsilon  (8 \epsilon +3 \cos3 \phi))^2)^{\frac{1}{2}}
~,
\end{eqnarray}
which, as before, strictly depends on the initial conditions of
the stellar encounter. A remark is in order at this point. A
monochromatic gravitational wave has, at most, two independent
degrees of freedom. In fact, in the TT gauge, we have  $h_{+} =
h_{11} + h_{22}$ and $h_{\times} = h_{12} + h_{21}$ (see e.g.
\cite{bla}). As an example, the amplitude of gravitational wave is
sketched in Fig. \ref{fig:ellisse800pc} for a stellar encounter
close to the Galactic Center. The adopted initial parameters are
typical of a close impact and are assumed to be $b=1$ AU  and
$v_{0}=200$ Km$s^{-1}$, respectively. Here, we have fixed
$M_{1}=M_{2}=1.4M_{\odot}$. The impact parameter is defined as
$L=bv$ where $L$ is the angular momentum and $v$ the incoming
velocity. We have chosen  a typical velocity of a star in the
galaxy and we are considering, essentially,  compact objects with
masses comparable to the Chandrasekhar limit $(\sim
1.4M_{\odot})$. This choice is motivated by the fact that
ground-based experiments like VIRGO or LIGO expect to detect
typical GW emissions from the dynamics of these objects or from
binary systems composed by them (see e.g. \cite{maggiore}).

\subsection{GW amplitude from parabolic and hyperbolic orbits}

In this case the single components of Eq.(\ref{ampli1}) for a
parabolic and hyperbolic orbits, are
\begin{equation}
\begin{array}{llllllll}
h^{11}=-\frac{G l^2 L^2\mu}{Rc^4}(13 \epsilon  \cos \phi+12 \cos2 \phi+\epsilon  (4 \epsilon +3 \cos3 \phi))
~,
\\ \\
h^{22}=\frac{Gl^2 L^2\mu}{Rc^4}(17 \epsilon  \cos\phi+12 \cos2 \phi+\epsilon  (8 \epsilon +3 \cos3 \phi))
 ~,
\\ \\
h^{12}=h^{21}=-\frac{3Gl^2 L^2\mu}{Rc^4} (4 \cos \phi+\epsilon  (\cos2 \phi+3)) \sin\phi~,
\end{array}
\end{equation}
and then the expected strain amplitude is

\begin{eqnarray}
h=\frac{2 l^4 L^4 \mu ^2 }{c^4 R}
 \left(10 \epsilon ^4+9  \epsilon ^3\cos 3 \phi+59 \epsilon ^2
   \cos2 \phi+59 \epsilon ^2+\left(47 \epsilon
   ^2+108\right)  \epsilon\cos\phi +36\right)^{\frac{1}{2}}
~,
\end{eqnarray}
which, as before, strictly depends on the initial conditions of
the stellar encounter. We note that the gravitational wave
amplitude  has the same analytical expression for both cases and
differs only for the value of $\epsilon$ which is $\epsilon=1$ if
the motion is  parabolic and the  polar angle range  is
$\phi\in(-\pi,\pi)$, while it is $\epsilon>1$ and
$\phi\in(-\pi,\pi)$ for hyperbolic orbits. In these cases, we have
non-returning objects.

\begin{figure}
\includegraphics[scale=0.6]{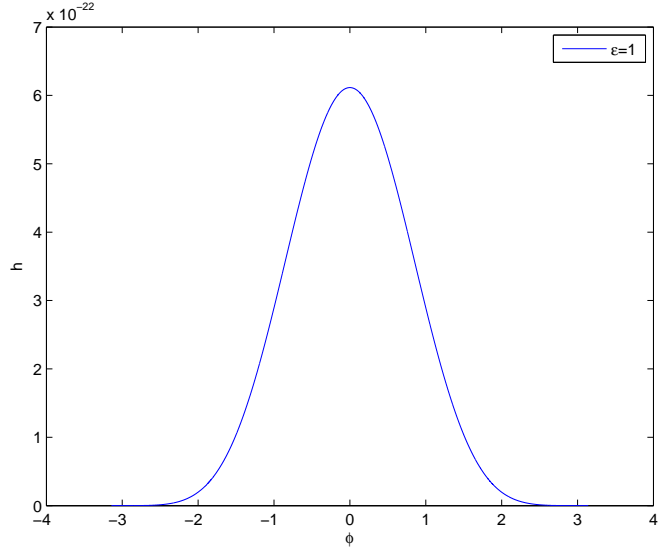}
\caption{The gravitational wave-forms for a parabolic encounter as
a function of the polar angle $\phi$.   As above,
$M_{1}=M_{2}=1.4M_{\odot}$ and $M_{2}$ is considered at rest.
$M_{1}$ is moving with initial velocity $v_{0}=200$ Km$s^{-1}$
with an impact parameter $b=1$ AU. The distance of the GW source
is assumed at $R=8$ kpc. The eccentricity is $\epsilon=1$. }
\label{fig:parabola}
\end{figure}

\begin{figure}
\includegraphics[scale=0.6]{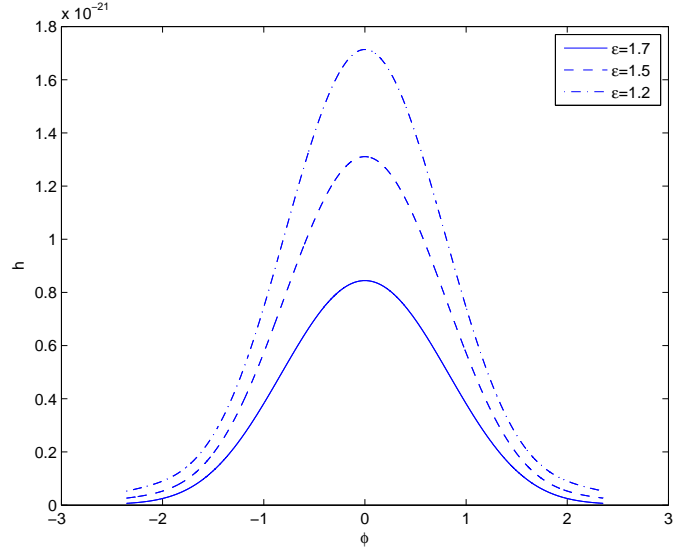}
\caption{The gravitational wave-forms for hyperbolic encounters as
function of the polar angle $\phi$.  As above, we have fixed
$M_{1}=M_{2}=1.4M_{\odot}$. $M_{2}$ is considered at rest while
$M_{1}$ is moving with initial velocity $v_{0}=200$ Km$s^{-1}$ and
an impact parameter $b=1$ AU. The distance of the source is
assumed at $R=8$ kpc. The eccentricity is assumed with the values
$\epsilon=1.2,1.5,1.7$ .} \label{fig:iperbole8}
\end{figure}

The amplitude of the gravitational wave is sketched in Figs.
\ref{fig:parabola} and \ref{fig:iperbole8} for  stellar encounters
close to the Galactic Center. As above, we consider  a close
impact and  assume  $b=1$ AU cm, $v_{0}=200$ Km$s^{-1}$ and
$M_{1}=M_{2}=1.4M_{\odot}$.

\section{Rate and event number estimations}

An important remark is due at this point. A galaxy is a
self-gravitating collisionless system where stellar impacts are
very rare \cite{binney}. From the GW emission point of view, close
orbital encounters, collisions and tidal interactions should be
dealt on average if we want to investigate the gravitational
radiation in a dense stellar system as we are going to do now.

Let us give now an estimate of the  stellar encounter rate
producing GWs in some interesting astrophysical conditions like a
typical globular cluster or towards the Galactic Center after we
have discussed above the features distinguishing the various types
of stellar encounters. Up to now, we have approximated stars as
point masses. However, in dense regions of stellar systems,  a
star can pass so close to another that they raise tidal forces
which dissipate their relative orbital kinetic energy. In some
cases, the loss of energy can be so large that stars form  binary
or multiple systems; in other cases, the stars collide and
coalesce into a single star; finally stars can exchange
gravitational interaction in non-returning encounters.

To investigate and parameterize all these effects, we have to
compute the collision time $t_{coll}$, where $1/t_{coll}$ is the
collision rate, that is, the average number of physical collisions
that a given star suffers per unit time. For the sake of
simplicity, we restrict  to stellar clusters in which all stars
have the same mass $m$.

Let us consider an encounter with initial relative velocity
$\mathbf{v}_{0}$ and impact parameter $b$. The angular momentum
per unit mass of the reduced particle is $L=bv_{0}$. At the
distance of closest approach, which we denote by $r_{coll}$, the
radial velocity must be zero, and hence the angular momentum is
$L=r_{coll}v_{max}$, where $v_{max}$ is the relative speed at
$r_{coll}$. From the energy equation (\ref{eq:energia}), we have

\begin{equation}
b^{2}=r_{coll}^{2}+\frac{4Gmr_{coll}}{v_{0}^{2}}\,.\label{eq:b}\end{equation}

If we set $r_{coll}$ equal to the sum of the radii of two stars,
then a collision will occur if and only if the impact parameter is
less than the value of $b$, as determined by Eq.(\ref{eq:b}).

Let $f(\mathbf{v}_{a})d^{3}\mathbf{v}_{a}$ be the number of stars
per unit volume with velocities in the range
$\mathbf{v}_{a}+d^{3}\mathbf{v}_{a}.$ The number of encounters per
unit time with impact parameter less than $b$ which are suffered
by a given star is just $f(\mathbf{v}_{a})d^{3}\mathbf{v}_{a}$
times the volume of the annulus with radius $b$ and length
$v_{0}$, that is,

\begin{equation}
\int f(\mathbf{v}_{a})\pi b^{2}v_{0}d^{3}\mathbf{v}_{a}\label{eq:integrale}\end{equation}

where $v_{0}=\left|\mathbf{v-v}_{a}\right|$ and $\mathbf{v}$ is
the velocity of the considered star. The quantity in
Eq.(\ref{eq:integrale}) is equal to $1/t_{coll}$ for a star with
velocity $\mathbf{v}$: to obtain the mean value of $1/t_{coll}$,
we average over $\mathbf{v}$ by multiplying (\ref{eq:integrale})
by $f(\mathbf{v})/\nu$, where $\nu=\int
f(\mathbf{v})d^{3}\mathbf{v}$ is the number density of stars and
the integration is over $d^{3}\mathbf{v}$. Thus
{}``\begin{equation}
\frac{1}{t_{coll}}=\frac{\nu}{8\pi^{2}\sigma^{6}}\int
e^{-(v^{2}+v_{a}^{2})/2\sigma^{2}}\left(r_{coll}\left|\mathbf{v-v}_{a}\right|+
\frac{4Gmr_{coll}}{\left|\mathbf{v-v}_{a}\right|}\right)d^{3}\mathbf{v}d^{3}\mathbf{v}_{a}
\label{eq:invtcoll}\end{equation}

We now replace the variable $\mathbf{v}_{a}$ by
$\mathbf{V}=\mathbf{v-v}_{a}$. The argument of the exponential is
then
$-\left[\left(\mathbf{v}-\frac{1}{2}\mathbf{V}\right)^{2}+\frac{1}{4}V^{2}\right]/\sigma^{2}$,
and if we replace the variable $\mathbf{v}$ by ${\displaystyle
\mathbf{v}_{cm}=\mathbf{v}-\frac{1}{2}\mathbf{V}}$ (the center of
mass velocity), then we have

\begin{equation}
\frac{1}{t_{coll}}=\frac{\nu}{8\pi^{2}\sigma^{6}} \int
e^{-(v_{cm}^{2}+V^{2})/2\sigma^{2}}\left(r_{coll}V+
\frac{4Gmr_{coll}}{V}\right)dV\,.\label{eq:invtcoll1}\end{equation}

The integral over $\mathbf{v}_{cm}$ is given by

\begin{equation}
\int
e^{-v_{cm}^{2}/\sigma^{2}}d^{3}\mathbf{v}_{cm}=\pi^{3/2}\sigma^{3}\,.\label{eq:intint}\end{equation}

Thus

\begin{equation}
\frac{1}{t_{coll}}=\frac{\pi^{1/2}\nu}{2\sigma^{3}}\int_{\infty}^{0}e^{-V^{2}/4\sigma^{2}}\left(r_{coll}^{2}V^{3}+4GmVr_{coll}\right)dV\label{eq:invtcoll2}\end{equation}

The integrals can be  easily calculated and then we find

\begin{equation}
\frac{1}{t_{coll}}=4\sqrt{\pi}\nu\sigma
r_{coll}^{2}+\frac{4\sqrt{\pi}\nu
Gmr_{coll}}{\sigma}\,.\label{eq:invtcooll3}\end{equation}

The first term of this result can be derived from the kinetic
theory. The rate of interaction is $\nu\Sigma\left\langle
V\right\rangle$, where $\Sigma$ is the cross-section and
$\left\langle V\right\rangle $ is the mean relative speed.
Substituting $\Sigma=\pi r_{coll}^{2}$ and $\left\langle
V\right\rangle =4\sigma/\sqrt{\pi}$ (which is appropriate for a
Maxwellian distribution whit dispersion $\sigma$) we recover the
first term of (\ref{eq:invtcooll3}). The second term represents
the enhancement in the collision rate by gravitational focusing,
that is, the deflection of trajectories by the gravitational
attraction of the two stars.

If $r_{*}$ is the stellar radius, we may set $r_{coll}=2r_{*}$. It
is convenient to introduce the escape speed from stellar surface,
${\displaystyle v_{*}=\sqrt{\frac{2Gm}{r_{*}}}}$, and to rewrite
Eq.(\ref{eq:invtcooll3}) as

\begin{equation}
\Gamma=\frac{1}{t_{coll}}=16\sqrt{\pi}\nu\sigma r_{*}^{2}\left(1+\frac{v_{*}^{2}}{4\sigma^{2}}\right)=16\sqrt{\pi}\nu\sigma r_{*}^{2}(1+\Theta),\label{eq:invtcoll4}\end{equation}

where

\begin{equation}
\Theta=\frac{v_{*}^{2}}{4\sigma^{2}}=\frac{Gm}{2\sigma^{2}r_{*}}\label{eq:safronov}\end{equation}

is the Safronov number \cite{binney}. In evaluating the rate, we
are considering only those  encounters producing gravitational
waves, for example,  in the LISA range, i.e. between $10^{-4}$ and
$10^{-2}$ Hz (see e.g. \cite{Rub}). Numerically, we have
\begin{equation}
\Gamma \simeq  5.5\times 10^{-10} \left(\frac{v}{10 {\rm km s^{-1}}}\right)
\left(\frac{\sigma}{UA^2}\right) \left(\frac{{\rm 10
pc}}{R}\right)^3 {\rm yrs^{-1}}\qquad\Theta<<1\label{eq:thetamin}
\end{equation}
\begin{equation}
\Gamma \simeq  5.5\times 10^{-10} \left(\frac{M}{10^5 {\rm
M_{\odot}}}\right)^2 \left(\frac{v}{10 {\rm km s^{-1}}}\right)
\left(\frac{\sigma}{UA^2}\right) \left(\frac{{\rm 10
pc}}{R}\right)^3 {\rm yrs^{-1}}\qquad\Theta>>1\label{eq:thetamagg}
\end{equation}

If $\Theta>>1$, the energy dissipated exceeds the relative kinetic
energy of the colliding stars, and the stars  coalesce into a
single star. This new star may, in turn, collide and merge with
other stars, thereby becoming very massive. As its mass increases,
the collision time is shorten and then there may be runaway
coalescence leading to the formation of a few supermassive objects
per clusters. If $\Theta<<1$, much of the mass in the colliding
stars may be liberated and forming new stars or a single
supermassive objects (see \cite{Belgeman,Shapiro}).

Note that when we have the effects of quasi-collisions in an
encounter of two stars in which the minimum separation is several
stellar radii, violent tides will raise on the surface of each
star. The energy that excites the tides comes from the relative
kinetic energy of the stars. This effect is important for
$\Theta>>1$ since the loss of small amount of kinetic energy may
leave the two stars with negative total energy, that is, as a
bounded  binary system. Successive peri-center passages will
dissipates more energy by GW radiation, until the binary orbit is
nearly circular with a negligible or null GW radiation emission.

Let us apply these considerations to the Galactic Center which can
be modelled as a system of several compact stellar clusters, some
of them similar to very compact globular clusters with high
emission in X-rays \cite{townes}.

For a typical \textbf{compact stellar cluster}  around the
Galactic Center, the expected event rate is of the order of
$2\times 10^{-9}$ yrs$^{-1}$ which may be increased at least by a
factor $\simeq 100$ if one considers the number of globular
clusters in the whole Galaxy eventually passing nearby the
Galactic Center. If the compact stellar cluster at the Galactic
Center is taken into account and assuming the total mass $M\simeq
3\times 10^6$ M$_{\odot}$, the velocity dispersion $\sigma\simeq $
150 km s$^{-1}$ and the radius of the object $R\simeq$ 10 pc
(where $\Theta=4.3$), one expects to have $\simeq 10^{-5}$ open
orbit encounters per year. On the other hand, if a cluster with
total mass $M\simeq 10^6$ M$_{\odot}$, $\sigma\simeq $ 150 km
s$^{-1}$ and $R\simeq$ 0.1 pc is considered, an event rate number
of the order of unity per year is obtained. These values could be
realistically achieved by data coming from the forthcoming space
interferometer LISA. As a secondary effect, the above  wave-forms
could constitute the "signature"  to classify the different
stellar encounters  thanks to the differences of the shapes (see
the above figures).

\section{Concluding Remarks}
We have analyzed the gravitational wave emission coming from
stellar encounters in Newtonian regime and in quadrupole
approximation. In particular, we have taken into account the
expected luminosity and the  strain amplitude of gravitational
radiation produced in tight impacts where two massive objects of
$1.4M_{\odot}$ closely interact at an impact distance of $1AU$.
Due to the high probability of such encounters inside rich stellar
fields (e.g. globular clusters, bulges of galaxies and so on), the
presented approach could highly contribute to enlarge the classes
of gravitational wave sources (in particular, of dynamical
phenomena capable of producing gravitational waves). In
particular, a detailed theory of stellar orbits could improve the
statistic of possible sources.

\begin{acknowledgments}
We  thank  G. Covone for  fruitful discussions  on the topics of
this paper. We thank also the Referee for the useful suggestions
which allowed to improve the paper.
\end{acknowledgments}

\end{document}